\documentclass[%
 reprint,
superscriptaddress,
 amsmath,amssymb,
 aps,
]{revtex4-2}

\usepackage{graphicx}% Include figure files
\usepackage{dcolumn}% Align table columns on decimal point
\usepackage{bm}% bold math

\usepackage[dvipsnames]{xcolor}
\usepackage{stackrel}

\usepackage{amsfonts}

\newcommand{\Dcrawl}{D_{\rm crawl}}
\newcommand{\Dhop}{D_{\rm hop}}
\newcommand{\pcrawl}{p_{\rm crawl}}
\newcommand{\phop}{p_{\rm hop}}
\newcommand{\degree}{^{\circ}}
\newcommand{\dx}{\Delta \text{x}}
\newcommand{\qon}{q_{\rm on}}
\newcommand{\qoff}{q_{\rm off}}
\newcommand{\Nbar}{N_b}
\newcommand{\tcrawl}{\tau_{\rm crawl}}
\newcommand{\thop}{\tau_{\rm hop}}
\newcommand{\tmode}{\tau_{\rm mode}}
\newcommand{\alphacrawl}{\alpha_{\rm crawl}}
\newcommand{\alphahop}{\alpha_{\rm hop}}
\newcommand{\alphamode}{\alpha_{\rm mode}}

\newcommand{\scrawl}{\sigma_{\rm crawl}}

\newcommand{\shop}{\sigma_{\rm hop}}

\usepackage[caption=false]{subfig}
\begin{document}

\preprint{APS/123-QED}

\title{Hopping and crawling DNA-coated colloids}

\author{Jeana Aojie Zheng}
\affiliation{Department of Physics, New York University, New York, NY 10003, USA}

\author{Miranda Holmes-Cerfon}
\affiliation{Department of Mathematics, University of British Columbia, Vancouver, BC, V6T 1Z2, Canada}

\author{David J. Pine}
\affiliation{Department of Physics, New York University, New York, NY 10003, USA}
\affiliation{Department of Chemical \& Biomolecular Engineering, New York University, New York, NY 11201, USA}

\author{Sophie Marbach}
\email{sophie.marbach@cnrs.fr}
\affiliation{Courant Institute of Mathematical Sciences, New York University, New York, NY 10012, USA}
\affiliation{CNRS, Sorbonne Universit\'{e}, Physicochimie des Electrolytes et Nanosyst\`{e}mes Interfaciaux, F-75005 Paris, France}

\date{\today}

\begin{abstract}
Understanding the motion of particles with ligand-receptors is important for biomedical applications and material design. %However, our understanding of such motion is still limited. 
%ability to control speed and mode of motion is impeded both by the multiple scales involved -- fast ligand binding over microscopic scales leads to macroscopic motion -- and by the diversity of mode of motions. 
Yet, even among a single design, the prototypical DNA-coated colloids, seemingly similar micrometric particles \textit{hop or roll}, depending on the study.
We shed light on this problem by observing DNA-coated colloids diffusing near surfaces coated with complementary strands for a wide array of coating designs. We find colloids rapidly switch between 2 modes: they \textit{hop} -- with long and fast steps -- \textit{and crawl} -- with short and slow steps. Both modes occur at all temperatures around the melting point and over a wide array of designs. The particles become increasingly subdiffusive as temperature decreases, in line with subsequent velocity steps becoming increasingly anti-correlated. Overall, crawling (or hopping) phases are more predominant at low (or high) temperatures; crawling is also more efficient at low temperatures than hopping to cover large distances. We rationalize this behavior within a simple model: at lower temperatures, the number of bound strands increases, and detachment of all bonds is unlikely, hence, hopping is prevented and crawling favored. We thus reveal the mechanism behind a common design rule relying on increased strand density for long-range self-assembly: dense strands on surfaces are required to enable crawling, possibly facilitating particle rearrangements. %Since dense strands on surfaces are required to enable crawling, we thus reveal the mechanism behind a common design rule: increased strand density enhances mobility and possibly improves ordered self-assembly. %Our work paves the way toward advanced and rational programming of ligand-receptor motion.

\end{abstract}

\maketitle

Understanding the motion of particles with ligand-receptors is of broad interest, for applications ranging from biomedical targeting~\cite{curk2017optimal,boitard2018magnetic,phan2023bimodal,phan2023bimodal,nerantzaki2022biotinylated,kowalewski2021multivalent} and screening~\cite{curk2020computational,xu2023whole} to material design~\cite{gehrels2022programming,rogers2016using} and water pollution remediation~\cite{talbot2021adsorption}. Such particles, a few nanometers to several microns in size, rely on specific binding and unbinding of up to thousands of fluctuating ligands---or \textit{feet}---to stick to receptor coated surfaces. Such ligand-receptor interactions also affect how the particles move, in ways that can be essential for their function. %For example, spherical hubs with protein-cleaving enzymes~\cite{unksov2022through,kowalewski2021multivalent} can mimick advanced molecular machines.
A prototypical example is DNA-coated colloids~\cite{rogers2016using}, which use DNA hybridization as the ligand-receptor bond that tethers colloids and mediates the self-assembly of large-scale colloidal crystals~\cite{he2020colloidal,mirkin1996dna,alivisatos1996organization,park2008dna,nykypanchuk2008dna,macfarlane2011nanoparticle,manoharan2016dnaccNatRevMat} or re-configurable colloidal molecules~\cite{chakraborty2022self,gehrels2022programming,mcmullen2022self}.
Assembly strongly depends on the relative motion of the DNA-coated colloids: if the bonds are too sticky, relative motion is limited, which prohibits particle rearrangements~\cite{geerts2010flying,holmes2016stochastic}. It is therefore crucial to understand the mechanisms governing relative motion of such DNA-coated surfaces.

%In fact,
%already understanding how such particles move remains an open question.
Nevertheless, mechanistic understanding of relative motion of such particles is still lacking. This is because of the complex, multiscale nature of the motion, with fast, small-scale, experimentally unresolvable ligand-receptor bonding dynamics giving rise to relative motion on the macroscale. This motion can take a variety of forms: ligand-receptor particles, can hop, roll, slide, crawl, glide or remain trapped~\cite{xu2011subdiffusion,sakai2017influenza,wang2015thermal,sakai2018unique,lowensohn2022sliding,wang2012brownian,geerts2010flying,jana2019translational}, depending on the microscopic bonding conditions. Even for a single well-defined system, the preferred mode of motion can vary: micron-sized DNA-coated colloids with similar coatings were seen to mostly hop~\cite{xu2011subdiffusion} or to perform cohesive moves~\cite{wang2015thermal}, to diffuse~\cite{hensley2022self,marbach2022nanocaterpillar,joshi2016kinetic,verweij2021conformations}, or to subdiffuse~\cite{xu2011subdiffusion,wang2015thermal}.
The multiscale nature of the motion challenges theoretical work~\cite{ding2014insights,angioletti2013communication,licata2007colloids,mitra2022coarse,lee2018modeling,marbach2022nanocaterpillar,jana2019translational,bartovs2023enhanced,janevs2022first,li2021biomechanics}, calling for high throughput experiments.

\begin{figure*}[htp!]
\centering
\includegraphics[width=\textwidth]{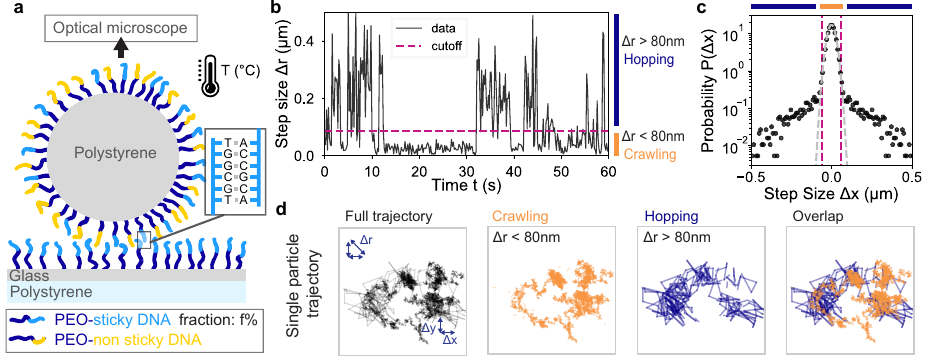}
\caption{\textbf{Intermittent hopping and crawling of a single DNA-coated colloid.} (a) Experimental setup. 1~$\mu$m-diameter polystyrene particles coated with DNA strands bind to and diffuse on a glass surface covered with complementary DNA. Their motion is tracked with an optical microscope. We tune the system's temperature and the fraction of sticky DNA on the particle. (b) Time series of step sizes for a single particle with $f = 5\%$ sticky ends around its melting temperature, at $T = 38.2^{\circ}$C and (c) step size distribution over the entire particle's trajectory. (d) Trajectory of (b), in black, colored, from left to right, as only steps smaller (orange) or larger (blue) than $82$~nm, and overlap. The trajectory displayed is $15~\mathrm{min}$ long and the box size is $2.8~\mathrm{\mu m}$.}
\label{fig:fig1}
\end{figure*}

In this work, we experimentally demonstrate that micron-sized DNA-coated colloids move on a DNA-coated surface alternating between 2 modes: \textit{hopping}, characterized by long steps in each resolvable time interval, and \textit{crawling}, characterized by short steps. This gives rise to a step-size distribution that is distinctly non-Gaussian, but well-modelled by a sum of two Gaussians with different widths. Both modes of motion are present over a range of temperatures spanning the melting point of the system, in single-particle trajectories and at the ensemble level. Hopping contributes the most to the mean-squared displacement of the particles at high temperatures, while crawling contributes the most at low temperatures.
%Particles switch between either mode more or less rapidly according to temperature and without memory.
 We build a theoretical model that reproduces most of the experimental features and brings mechanistic insight: at high temperatures, the number of bonds with the surface is small enough to permit coordinated detachment of all bonds, diffusion in free space, then reattachment, corresponding to the long steps observed in hopping. At lower temperatures, the number of bonds is higher and hence a cohesive mode of motion dominates, akin to crawling. Finally, we shed light on the mechanism yielding subdiffusion at low temperatures: anti-correlation between subsequent steps determines the subdiffusive exponent. %This mechanistic insight is confirmed experimentally as the bound to unbound fraction of colloids closely follows the probability to hop.
 Our work thus bridges an existing literature gap, showing micron-sized DNA-coated colloids may \textit{hop} \textit{and} \textit{crawl} on a surface, %, with one mode or the other dominating according to physical parameters. Our experimental approach 
 and paves the way towards rational programming of ligand-receptor mediated processes.

\noindent \textit{\textbf{DNA-coated colloids hop and crawl at the single particle level.}}
We track with an optical microscope the motion of 800 DNA-coated colloids, $R = 500~$nm in radius, as they diffuse on a DNA-coated substrate (Fig.~\ref{fig:fig1}-a). Our fabrication procedure is similar to previous work~\cite{oh2015high,cui2022comprehensive} (SM Sec. 1). Briefly, on the polystyrene particle, single stranded DNA is anchored through a polyethelyne oxide linker %with molecular weight 6500 g/mol 
using click-chemistry~\cite{agard2004strain}. The DNA strand is 20 nucleotides long (IDT, Coralville, IA), including a 14-poly-T tether followed by a 6-nucleotide ``sticky end" that can hybridize with complimentary strands on the substrate. The brush-mediated DNA functionalization results in a high-density DNA coating of about $0.1~\mathrm{nm^{-2}}$. We can modulate the fraction of sticky ends on the brush, between $5-100\%$. Here, we show results for $5~\%$ sticky ends, but we find qualitatively similar results for all other sticky fractions explored, corresponding to a range of temperatures $25-65^{\circ}C$ (SM Sec. 6). 
At each temperature, particles are tracked for about $20~\mathrm{min}$ at $\Delta t = 0.2~\mathrm{s}$ intervals. Images are then analyzed using the TrackPy software to obtain individual particle positions $x(t)$ and $y(t)$ along the surface with time $t$~\cite{crocker1996methods}. 

A single particle already demonstrates 2 types of mobility. Fig.~\ref{fig:fig1}-b shows a time series of the magnitude of the displacement, $\Delta r=\sqrt{\Delta x ^2 + \Delta y^2}$ where $\Delta x, \Delta y$ are horizontal displacements undergone by the particle in between each frame.
The particle alternates between taking many short steps, punctuated by bursts of longer steps.
A histogram of step sizes $\dx$, representing both the horizontal $ \Delta x$ and vertical $\Delta y$ increments of a particle's trajectory, is distinctly non-Gaussian (Fig.~\ref{fig:fig1}-c), with a sharp kink at the transition to a heavier-tailed region.
We use this kink to define a step-size cutoff distinguishing short and long steps -- which determines the placement of the dashed pink line in Fig.~\ref{fig:fig4}-b -- and then color the particle's 2D trajectory (Fig.~\ref{fig:fig1}-d) according to whether the steps are short (orange) or long (blue).
This gives another picture of the particle's motion, showing that short steps correspond to motion that is fairly localized, where exploration is limited---which we refer to as \textit{crawling}---whereas long steps allow the particle to move to farther regions---which we refer to as \textit{hopping}. %Furthermore, the distribution of steps averaged over many particles shows a clear deviation from a Gaussian, supporting this preliminary approach (Fig.~\ref{fig:fig1}-c).
These 2 modes of motion occur everywhere on the sample and can be observed repeatedly for particles whose trajectories are long enough (SM Sec.~2). %, showing that dual-mobility is an intrinsic feature of DNA-coated colloids.

\begin{figure}[htp!]
\centering
\includegraphics[width=\linewidth]{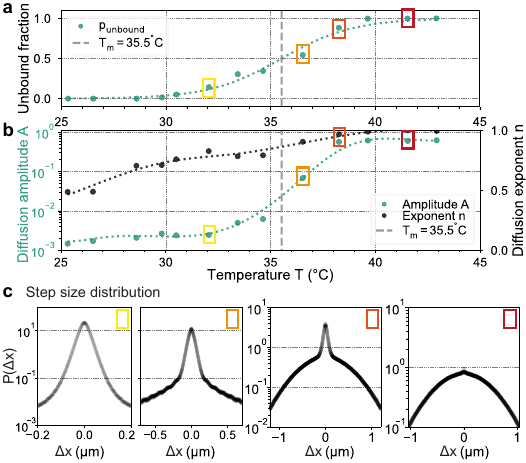}
\caption{\textbf{Step sizes change drastically with temperature.} (a) Fraction of unbound particles with $f = 5\%$ sticky ends as in Fig.~\ref{fig:fig1}. (b) Diffusion amplitude $A$ (yellow) characterizing the slow-down with temperature and diffusion exponent $n$ (gray, right axis). Dotted lines in (a-b) are guides to the eye. (c) Step size distribution over about 800 particle trajectories at increasing temperatures ($T = 32.1,36.5,38.3,41.5^{\degree}$C) marked by the colored boxes (yellow to red). The 2 left-ward plots share the same vertical axis.}
\label{fig:fig2}
\end{figure}

\noindent \textit{\textbf{Diffusion, crawling and hopping properties depend on temperature.}}
To gain more insight into the particles' mobility, we investigate the motion %measure the mean-squared displacement $\langle r^2(t) \rangle$ 
of the particles at different temperatures. The fraction of unbound particles $p_{\rm unbound}$, which we define by the number of particles that go out of focus for at least $1~\mathrm{min}$, increases sharply around a critical temperature $T_m$, the melting temperature~\cite{rogers2016using} (Fig.~\ref{fig:fig2}-a). 
We measure the mean-squared displacement $\langle r^2(t) \rangle$ at each temperature, averaging over all particles and over each trajectory. We then fit the data, using a standard least-squares procedure, as $\langle r^2(t) \rangle  = 4 A D_0 t_0 (t/t_0)^{n}$.
Here $D_0 = k_B T/6 \pi \eta R \simeq 1~\mathrm{\mu m^2/s}$ is the bulk diffusion coefficient of the particle (with $k_B T$ the thermal energy, $R$ the particle radius, and $\eta$ the fluid viscosity), $t_0 = R^2/4D_0 \simeq 63~\mathrm{ms}$ is the time for the particle to diffuse its diameter, and we fit for $A$ and $n$ (Fig.~\ref{fig:fig2}-b).
The particle's motion is diffusive ($n \simeq 1$) at high temperatures, with a diffusion amplitude $A \simeq 0.5$, corresponding to increased hydrodynamic friction near the substrate~\cite{sprinkle2020driven,goldman1967slow}.
Since the depth of focus is roughly the size of the particles $\sim 560~\mathrm{nm}$, we may estimate that imaged particles are $10-50~\mathrm{nm}$ from the surface. Such distances yield a hydrodynamic diffusion amplitude $A$ that varies only weakly with distance to the surface~\cite{goldman1967slow,lavaud2021stochastic,sprinkle2020driven}.
Using the logarithmic scaling law in Ref.~\cite{sprinkle2020driven} gives that $A^{\rm (th)} \simeq 0.53-0.67$, where the superscript $^{(th)}$ indicates theory predictions, and is close to the experimental value. As the temperature decreases, motion progressively becomes subdiffusive ($n < 1$), especially at low temperatures $T \simeq T_m - 10\degree$C where $n \simeq 0.5$. This property was already highlighted in previous work~\cite{xu2011subdiffusion,wang2015thermal}. Concomitantly, the diffusion amplitude radically slows down around the melting temperature, where $A$ decreases by about 3 orders of magnitude. Our goal is to understand how the 2 modes of motion are related to this dramatic decrease in diffusion amplitude and the subdiffusive behavior. 

The step size distribution $P(\dx)$ of all particles drastically changes with temperature (Fig.~\ref{fig:fig2}-c). At high temperatures, when particles are unbound, the distribution is close to a single Gaussian distribution (Fig.~\ref{fig:fig2}-c, right, red box). %, corresponding to free diffusion near the surface.
At a slightly lower temperature (Fig.~\ref{fig:fig2}-c, right, orange box) a central peak emerges in the step size distribution, an indicator of dual-mobility. %Since the center peak is narrow, it corresponds to a slower mode of motion, crawling.
As temperature decreases further this peak becomes predominant (Fig.~\ref{fig:fig2}-c left, orange and yellow boxes). The temperature dependence of the distribution rules out that non-Gaussianity could be caused by local hydrodynamic friction close to the surface~\cite{matse2017test,lavaud2021stochastic,chechkin2017brownian}. The non-Gaussian distributions may therefore be attributed to multiple mobility modes~\cite{wang2012brownian}.

\begin{figure}[htp]
\centering
\includegraphics[width=\linewidth]{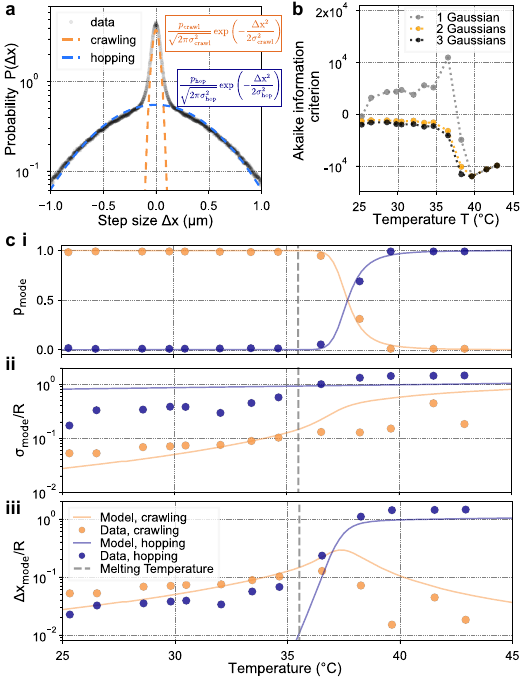}
\caption{ \textbf{Various modes of motion according to temperature} (a) Example of 2-Gaussians fit procedure on a representative data set for $f = 5\%$ sticky ends at $T = 38.3^{\degree}$C, see more examples in SM Sec.~3. (b) Akaike information criterion for the 1, 2 or 3-Gaussians fitting procedure with temperature for $f = 5\%$ sticky ends. (c) Extracted fitted parameters, from top to bottom: (i) probability, (ii) typical step size and (iii) efficiency in either mode; with temperature for $f = 5\%$ sticky ends. The legend is common to (i-iii). Fig.~S12 reports (i) in log-scale on the y-axis.
}
\label{fig:fig3}
\end{figure}

\begin{figure}[hbp]
\centering
\includegraphics[width=\linewidth]{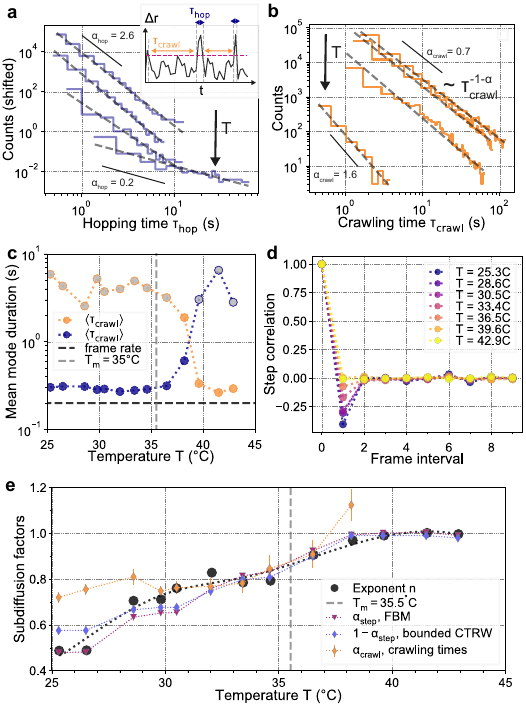}
\caption{\textbf{Subdiffusion explained by non-Markovian signatures in both mobility modes} (a) Distributions of hopping and (b) crawling mode durations, as illustrated in the inset of (a). The dashed lines are power-law fits as $\tau^{-1 - \alpha}$. Here, $T = 25.3, 32, 36.5, 39.6 \degree$C, with increasing temperatures presented from top to bottom. In (a) only, each curve is displaced by a factor 10 to enhance visibility, with the highest curve having actual units. $f= 5\%$ sticky ends on the particle. (c) Mean mode duration with temperature as obtained from averaging experimental data; notice the log scale on the y-axis. Gray points correspond to points where the average is ill-defined since $\alpha <1$. (d) Step correlation as a function of the interval between frames at different temperatures. (e) Subdiffusion factors obtained from (b) or (d), see text for details, compared to the subdiffusion exponent $n$. Dotted lines are guides for the eyes. The data for $\alpha_{\rm crawl}$ only goes up until $T = 38.2 \degree$C since beyond that point, data is not statistically relevant to fit power law distributions.
}
\label{fig:fig4}
\end{figure}

To unravel the temperature-dependent properties of each mode, we fit the step size distributions with a sum of 2 zero-mean Gaussians with different widths (Fig.~\ref{fig:fig3}c),
\begin{equation}
\begin{split}
    P(\dx) = &\frac{\phop}{\sqrt{2\pi \shop^2}} \exp\left( - \frac{\dx^2}{2\shop^2 }  \right)  \\
    &+ \frac{\pcrawl}{\sqrt{2\pi \scrawl^2}} \exp\left( - \frac{\dx^2}{2 \scrawl^2} \right).
\end{split}
 \end{equation}
 The parameters have a natural interpretation: $\phop  = 1 - \pcrawl$ is the probability to hop and $\pcrawl$ to crawl, and $\scrawl$ and $\shop$ are the characteristic step sizes in either mode. 
We use a least-squares procedure to  fit for $\phop$, $\shop$, and $\scrawl$. The Akaike information criterion (AIC) informs on the likelihood that this fit is representative of the data~\cite{cavanaugh2019akaike}. The AIC for this 2-Gaussian fit is much smaller than that for a 1-Gaussian fit at low temperatures (Fig.~\ref{fig:fig3}-b), indicating that 2 Gaussians provide a better characterization. Adding a  3rd Gaussian barely improves the AIC, confirming that 2 Gaussians is the most informative model. At high temperatures, the step size distributions approach a single Gaussian, consistent with the expectation that at these temperatures, colloids should move freely. Our approach is not sensitive to the fitting procedure (SM Sec.~3). %We thus obtain the probability to be in either mode, $\phop$, $\pcrawl$ and the typical step size $\shop$ and $\scrawl$ (dots in Fig.~\ref{fig:fig3}-c), with respect to temperature.

The extracted probabilities to be in either mode, $\phop, \pcrawl$, depend strongly on temperature (Fig.~\ref{fig:fig3}-c.i).
These probabilities undergo a sharp transition a few degrees above $T_m$, with crawling (or hopping) being more likely below (or above) $T_m$.
This can be understood in the light of the melting curve in Fig.~\ref{fig:fig2}-a. At low temperatures, ligands are more likely to form bonds with the surface receptors, thereby slowing the particles' motion. %, and further justifying the use of the word \textit{crawling}.
We further observe that both characteristic step sizes $\shop$ and $\scrawl$ decrease as temperature is lowered (Fig.~\ref{fig:fig3}-c.ii). This is again consistent with the melting curve of Fig.~\ref{fig:fig2}-a, since at lower temperatures we expect more ligand-receptor bonds, further inhibiting motion. %In addition, cooling brings the particle closer to the surface~\cite{cui2022comprehensive}, possibly increasing hydrodynamic friction further \cite{sprinkle2020driven}.

Which mode contributes the most to the particles' overall mobility?
The mean squared displacement in one time step according to our fitting model is $\langle \dx^2 \rangle =  \pcrawl \scrawl^2 + \phop \shop^2$. Therefore, we may define $\dx_{\rm crawl} = \sqrt{\pcrawl}\scrawl$ and $\dx_{\rm hop} = \sqrt{\phop}\shop$ to be the \textit{effective} distance covered by either crawling or sliding in one step. Even though hopping steps are longer than crawling ones, $\shop \gg \scrawl$, a particle can still cover more territory by crawling if the probability to hop $\phop$ is small. 
We find crawling is slightly more efficient at low temperatures, $\dx_{\rm crawl} \gtrsim \dx_{\rm hop}$, whereas hopping is more efficient above the melting temperature, $\dx_{\rm hop} \gg \dx_{\rm crawl}$ (Fig.~\ref{fig:fig3}-c.iii).

\noindent \textit{\textbf{A microscopic model determines crawling is predominant at low temperatures because it does not require to detach numerous bonds.}}
We rationalize our observations using a previously introduced model of DNA-coated colloid motion mediated by rapid binding and unbinding of the DNA strands (SM Sec.~4)~\cite{marbach2022nanocaterpillar}.  We first determine the number of nearby strands available for binding $N$ and the average number of bonds $\Nbar$ with temperature, accounting for steric and hybridization interactions~\cite{cui2022comprehensive,angioletti2013communication,santalucia1998unified}. We then add kinetics to our model, by assuming strands bind independently with binding and unbinding rates $\qon$ and $\qoff$ respectively~\cite{marbach2022nanocaterpillar}, which are related via the average number $\Nbar = N \frac{\qon}{\qon + \qoff}$. Once bound, each strand exerts a recoil force on the particle, modeled by a harmonic spring force with spring constant $k$ based on a worm-like chain model for the polymers~\cite{marbach2022nanocaterpillar}. In addition, we include the possibility for colloids to diffuse far from the surface, reporting details in SM Sec. 4.1. 
%~\cite{martinez2014designing}.
Coarse-graining over the fast strand motion and binding kinetics, we obtain an analytic expression for the effective diffusion coefficient of the particle as a function of the microscopic DNA ($N$, $\qon$, $\qoff$, $k$, and strand length) and particle parameters (its radius $R$). All these parameters are known from experimental data, except the density of strands on the glass surfaces, which is fitted once to obtain the correct melting temperature and is comparable ($0.009~\mathrm{nm^{-2}}$) with previous work~\cite{cui2022comprehensive,marbach2022nanocaterpillar}.

Within our model, the probability of hopping is the equilibrium probability that no bonds are formed,
\begin{equation}
    \phop^{\rm (th)} = \left( \frac{\qoff}{\qoff + \qon}\right)^N.
\end{equation}
This probability depends on $T$, since as temperature decreases, %sticky surfaces get closer, which increases the number of accessible partners $N$~\cite{cui2022comprehensive}, and decreases 
the DNA hybridization energy decreases~\cite{santalucia1998unified} so that $\qoff$ should decrease. We expect $\qon$ to be roughly constant with temperature~\cite{zhang2018predicting}.
The model remarkably captures  $\phop$ and $\pcrawl$ around the melting temperature, and thus the transition between hopping and crawling (Fig.~\ref{fig:fig3}-c.i), supporting our hypothesis that hopping arises when all bonds simultaneously detach.

The model's prediction for the mean step size in each mode is obtained by assuming that motion in each mode is Brownian (steps are gaussian and uncorrelated), so that $(\sigma^{\rm (th)}_{{\rm mode}})^2 = 2 D^{\rm (th)}_{\rm mode} \Delta t$, where $D^{\rm (th)}_{\rm mode}$ are the diffusion coefficients associated with each mode, given by our theory~\cite{marbach2022nanocaterpillar}
%The diffusion coefficient for crawling scales inversely with the average number of bonds $\Nbar = N \frac{\qon}{\qon + \qoff}$,
as
\begin{equation}
    \Dcrawl^{\rm (th)} \simeq \displaystyle \frac{D^{\rm hydro}_0}{1 + \Nbar \frac{k}{\qoff} \frac{k_B T}{D^{\rm hydro}_0}}, \qquad
    \Dhop^{\rm (th)} = D^{\rm hydro}_0.
\end{equation}
Here $D^{\rm hydro}_0=A^{\rm (th)}D_0$ is the diffusion coefficient of the unbound particle, which accounts for increased hydrodynamic friction with the substrate~\cite{sprinkle2020driven}.
The approximate expression for $\Dcrawl^{\rm (th)}$ is valid for large $N$.
The measured and theoretically predicted effective step sizes in Fig.~\ref{fig:fig3}-c.ii agree reasonably well at some but not all temperatures. The measured and theoretically predicted efficiency of each mode in Fig.~\ref{fig:fig3}-c.iii agree when both the step size and probability agree.

The regions where the theory and measurements agree inform us about the microscopic mechanisms underlying the mode of motion.  
%We delve into the potential mechanisms uncovered by the theory, before discussing discrepancies between theory and data. 
The effective step sizes for crawling, $\sigma_{\rm crawl},\sigma^{\rm (th)}_{\rm crawl}$ (orange markers / line) agree well for $T<T_m$, both decreasing with temperature (Fig.~\ref{fig:fig3}-c.ii).
The decrease in the theoretical step size arises because the average number of bonds $\Nbar$ increases, which increases the effective friction on the particle. An increase in the number of bonds is thus a potential mechanism for the increasingly shorter steps at low temperatures.
The temperature at which both modes are equally efficient, both experimentally and theoretically, occurs slightly above the melting temperature and also corresponds to a maximum in the crawling efficiency (Fig.~\ref{fig:fig3}-c.iii). The maximum corresponds to a trade-off between less probable yet longer crawling steps with increasing temperatures. The optimal crawling speed in our model corresponds roughly with $N_b \simeq 4$ bonds (note that $N_b (T_m) = 9$). This optimal bond value for mobility resonates with Ref.~\cite{martinez2014designing}, where a similar trade-off was predicted in a simulation of particles walking up concentration gradients %: again faster speed is achieved with a lower amount of bonds since bonds exert recoil forces on the particle, yet more bonds drive a particle faster up concentration gradients, 
and the optimal bond number was around $N_b \simeq 5$. %This sparks the idea that universal principles may govern DNA-coated colloid mobility.

Discrepancies between theory and data also shed further insight on the dual hopping-crawling motion.
For $T > T_m$, the theory over-predicts the measured steps, $\sigma^{\rm (th)}_{\rm crawl}>\sigma_{\rm crawl}$. We speculate that crawling steps may be dominated by outlier particles with a slightly higher local ligand or receptor density. These are not accounted for in our mean-field model, and such corrections of the model are also needed to explain a broader experimental melting curve than predicted by the theory (SM 4.2). %Outliers with slightly higher density locally on their surface, may likely still crawl with a few strands attached.
The effective step sizes for hopping, $\sigma_{\rm hop}$ and $\sigma^{\rm (th)}_{\rm hop}$ (blue markers/line) agree within 15\% at high temperatures, $T>T_m$, with a slight mismatch possibly attributable to variability in the exact particle size or density, with slightly smaller particles diffusing faster and lighter ones further away from the surface having less hydrodynamic friction~\cite{cui2022comprehensive}. 

At low temperatures, $T<T_m$, the model overpredicts $\sigma_{\rm hop}$ by a factor of 2-7.
%
%We will later argue this overprediction arises because we cannot resolve the hopping steps at low temperatures, hence the measurements of hopping are actually a mix of hopping and crawling steps. However, let us first explore some alternative reasons for the mismatch.
There are several possible reasons for this discrepancy.
One possibility is the increased hydrodynamic friction from the soft polymer mesh~\cite{hill2003electrophoresis,bertin2022soft}. Using Brinkman lengths and polymer brush thicknesses obtained in a previous work~\cite{cui2022comprehensive}, we find this amounts to decreasing $\sigma_{\rm hop}$ by about a factor 2, not enough to explain the discrepancy.
%Lateral hydrodynamic friction on the hard surfaces scales logarithmically with the distance~\cite{sprinkle2020driven,lavaud2021stochastic}, not enough to explain the discrepancy.
Another possibility could be connected with the subdiffusive nature of motion for temperatures below the melting temperature. Assuming the actual displacement is $\shop^{\rm (n, th)} = \shop^{\rm (th)} (\Delta t/t_0)^{(n-1)/2}$ where $n$ is the experimentally measured subdiffusive exponent in Fig.~\ref{fig:fig2}-b. Taking $n=0.5$ gives $\shop^{\rm (n, th)} \simeq 0.7 \shop^{\rm (th)}$ which is also not sufficient to explain the discrepancy.
A third possibility is that we are not resolving the hopping steps at low temperatures, hence the measurements of hopping are actually a mix of hopping and crawling steps. We explore this possibility next.

\noindent \textit{\textbf{Switching times are power-law distributed, but this does not relate to subdiffusion.}}
To investigate if we can resolve the different modes in time at all temperatures, we extracted the durations of crawling and hopping events from our data, $\tcrawl$, $\thop$ (Fig.~\ref{fig:fig4}), by determining a step size cutoff from the intersection point between the 2-Gaussian fit at each temperature. The average duration of each mode, $\langle \tcrawl \rangle$ and $\langle \thop \rangle$,  dramatically changes with temperature (Fig.~\ref{fig:fig4}-a). Crawling phases last about $\langle \tcrawl \rangle \simeq 8~\mathrm{s}$ at low temperatures but are much shorter at high temperatures, $\langle \tcrawl \rangle \simeq \Delta t$.
In contrast, hopping phases are short at low temperatures, $\langle \thop \rangle \simeq \Delta t$, but longer at high temperatures,  $\langle \thop \rangle \simeq 3~\mathrm{s}$.
These observations suggest that we are not fully resolving hopping phases at low temperatures and crawling phases at high temperatures. Rather, steps that we identify as ``hopping'' at low temperatures, may actually contain a mixture of hopping and crawling motions, making the measured step size smaller than it would otherwise be. This is consistent with our theoretical model since the model overestimated hopping step sizes at low temperatures. It is also consistent with SI Fig.~S12 showing probabilities of either mode in log scale (Fig.~\ref{fig:fig3}-c.iii in log scale), and where we find experimental probabilities to hop are higher than in the model. Similarly, steps that we identify as ``crawling'' at high temperatures, also contain some hopping, making them longer than they otherwise would be. Yet, at high temperatures, the model overestimated crawling step sizes. Differences between the model and experiments at high temperatures may be dominated by outlier particles with a higher ligand or receptor density.

The distributions of durations can be clearly accounted for with power laws as $p(\tmode) \simeq \tmode^{-1 - \alphamode}$ (Fig.~\ref{fig:fig4}-a and b), at all temperatures (see also SM Sec. 5.1). The power-law scaling is a sign of non-Markovian dynamics, or memory, since switching times should be exponentially distributed in a Markovian process. Previous work that investigated the distribution of times that particles appeared to be bound have also observed similar power laws~\cite{rogers2013kinetics,biancaniello2005colloidal,xu2011subdiffusion}, which can be related to our distribution of crawling times. The exponent of the power law $\alphamode$ depends on the mode and on temperature $T$. For hopping steps, we find $\alphahop \simeq 2-3$ at low temperatures while $\alphahop \simeq 0.2$ at high temperatures. In contrast, for crawling steps, $\alphacrawl \simeq 0.7 - 0.9$ at low temperatures and increases at high temperatures up to $\alphacrawl \simeq 1.5-2$. When $\alpha < 1$, the power law distribution predicts that the mean duration of a mode diverges $\langle \tmode \rangle = \infty$, since $\int_0^{\infty} t (t^{-1-\alpha}) \mathrm{d} t$ diverges. We highlight the temperature points where $\alphamode <1$ as gray dots in Fig.~\ref{fig:fig4}-c, indicating that experimental estimates of $\langle \tmode \rangle$ are likely inaccurate for these points. In practice, the duration of a mode may still be finite, since long tails, including for DNA-coated colloid binding, can have exponential decays in the very long range~\cite{rogers2013kinetics,biancaniello2005colloidal}. Overall, the particles thus spend extremely long times in the crawling or hopping modes at low or high temperatures, respectively. 

How are these slow modes connected to subdiffusion? A previous investigation of only-hopping DNA-coated colloids~\cite{xu2011subdiffusion} highlighted that bound time distributions with infinite tails could be connected to subdiffusion. This is supported even more by the fact that, here, the exponent characterizing subdiffusion $n$, as $\langle r^2(t) \rangle \sim t^n$,  (Fig.~\ref{fig:fig4}-e, black) appears to be in close agreement, at least for $T \gtrsim 30^{\circ}C$, with the exponent for the power law distributions $n \simeq \alphacrawl$ (orange diamonds). However, we could find no explanation for this fortuitous agreement. Models of continuous time random walks (CTRW) assume jumps at random time intervals, distributed with long-tailed time distributions in between jumps. Yet, for CTRW, the mean-squared displacement, averaged over particles and initial times, scales linearly as $\langle r^2(t) \rangle \sim t$, and it is a common misconception that CTRW should result in subdiffusion~\cite{burov2011single,rehfeldt2023random,ernst2012fractional,he2008random}. In addition, our system here is more complex than a standard CTRW, since we alternate between 2 mobile modes.

\noindent  \textit{\textbf{Anticorrelation between subsequent steps is related to subdiffusion.}}
To unravel alternative connections between subdiffusion and our mobility modes, we calculate the velocity step correlation between two consecutive jumps $\langle \dx(t) \dx(t+ n_f\Delta t) \rangle$ with a variable number of frames $n_f$ in between steps (Fig.~\ref{fig:fig4}-d, $n_f=1$). Since it is hard to disentangle crawling and hopping at some temperatures, we combine all modes for this analysis. At low temperatures, we find the steps are anti-correlated. The magnitude of the anticorrelation increases with decreasing temperatures and is independent of the number of frames $n_f$ between steps (SM 5.2).
%The anticorrelation peak occurs at a frame interval of $n_f = 1$, but does not depend on the frame rate~\cite{burov2011single}, meaning that the anticorrelation behavior's time scale is shorter than 0.2~s. 
Such a signature of non-Markovianity, or memory, is reminiscent of both Fractional Brownian Motion (FBM) and a bounded CTRW (bCTRW)~\cite{burov2011single}. To compare to FBM, we compute the magnitude of the anticorrelation peak as $a_1 = \langle \dx(t) \dx(t+ \Delta t) \rangle/ \langle \dx(t)^2 \rangle$. For an FBM, this magnitude is 
$a_1 = 2^{\alpha_{\rm FBM} - 1}$, where $\alpha_{\rm FBM}$ is the subdiffusive exponent,  $\langle r^2(t) \rangle \sim t^{\alpha_{\rm FBM}}$ (and the magnitude is independent of the frame rate)~\cite{burov2011single}. We find remarkable agreement between $\alpha_{\rm FBM}$ (Fig.~\ref{fig:fig4}-e, purple triangles) and the subdiffusive exponent $n$ (black dots), at all temperatures. Similarly, one can compare to a bounded CTRW (bCTRW), arguing that dense patches of sticky DNA strands may be at the origin of the confinement. In that case, the magnitude of the anticorrelation peak is related to an exponent $\alpha_{bCTRW}$, through a complex formula not reported here~\cite{burov2011single}, which predicts $\langle r^2(t) \rangle \sim t^{1-\alpha_{\rm bCTRW}}$. The exponent $1-\alpha_{\rm bCTRW}$ (Fig.~\ref{fig:fig4}-e, blue diamonds) is also in good agreement with $n$. 
Determining whether our system is better described by a FBM or a bCTRW requires further analysis such as ergodicity or asphericity measurements~\cite{rehfeldt2023random,ernst2012fractional}. Unfortunately, here our data sample sizes are too small to yield significant results and to discriminate between these categories. However, 
this highlights that the origin of subdiffusion may lie in subsequent anti-correlated steps. Whether these correlations are more due to hopping or crawling is still an open question to be resolved with higher frame rates and/or 3D investigations.

\noindent \textit{\textbf{Discussion.}}
In summary, we have observed 2 \textit{simultaneous} mobility modes for micron-sized DNA-coated colloids: \textit{hopping}, corresponding to fast and long steps, that dominates at temperatures above the melting temperature; and \textit{crawling} with slow and short steps that dominates below. Both hopping and crawling occur at the single particle level as well as the ensemble level, and a single particle switches rapidly between the 2 modes, with power-law distributed switching times.
Within our theoretical model that captures the main features of the experiments, we interpret that hopping corresponds to events where a particle detaches all bonds from the surface, floats in free space, and reattaches, while crawling is a cohesive move on the surface, where the particle is always in contact through a few bonds.
Crawling slows down with decreasing temperatures, as more bonds form and exert recoil forces -- explaining the strong mobility slow down by orders of magnitude. The time intervals spent crawling are apparently divergent at low temperatures. At low temperatures, consecutive steps are significantly anti-correlated, consistent with classical random walk models with memory. These models highlight that these anti-correlated steps are consistent with the overall subdiffusive motion of the particles.

Our analysis sheds light on seemingly disparate results of mobility in the literature~\cite{wang2015thermal,xu2011subdiffusion}. In fact, the DNA-coated colloids in Ref.~\cite{xu2011subdiffusion} were likely only observed to hop, because the coatings were low density, preventing strands from extending in a coordinated fashion to crawl. Since hopping efficiency decays even faster with cooling than crawling, this explains the extremely fast slowdown of the colloids in Ref.~\cite{xu2011subdiffusion}. In contrast, the DNA-coated colloids in Ref.~\cite{wang2015thermal} were only observed to perform some cohesive mode of motion at low temperatures. Since they are densely coated, similar to ours, and were only investigated below the melting temperature, our analysis shows cohesive motion is predominant. Since we find both hopping and crawling occur systematically for DNA-coated colloids with various coatings, we hypothesize that micron-sized DNA-coated colloids can \textit{hop and crawl} in general. As we have seen, either mode is not necessarily as efficient, depending on the temperature and the coating design. Yet, the ability to crawl improves mobility at low temperatures and hence may explain why high-density colloids might be preferred for improved self-assembly~\cite{cui2022comprehensive,geerts2010flying}.

In addition, we expand our understanding of the mechanisms underlying subdiffusion of DNA-coated colloids. Previous results on low-density coatings in Ref.~\cite{xu2011subdiffusion} found colloids would remain stuck for very long times in between subsequent hops. The waiting times were distributed with power-laws, with exponents $\alpha < 1$, which is consistent with our measured distributions of crawling times (Fig.~\ref{fig:fig4}-b). However, we find that this is not sufficient to explain subdiffusion at temperatures well below the melting temperature, and that, in addition, crawling and/or hopping steps are anti-correlated. A particle is, therefore, more likely to come back to its original position when it attempts to move away. Unfortunately, Ref.~\cite{xu2011subdiffusion} does not report an investigation of correlations between subsequent steps. %We, therefore, hope to motivate future research to perform such an analysis. 
The origin of such anti-correlation must lie in the microscopic details of binding and unbinding, which is hard to relate to macroscopic behavior~\cite{marbach2022nanocaterpillar,lee2018modeling} and we hope to motivate further research in that field as well.  

Since our method is limited to 2D tracking, it is not possible to distinguish here between various potential crawling modes: for example between sliding or rolling DNA-coated colloids. With the advent of 3D, super-resolved, microscopy techniques as well as high-throughput techniques to synthesize designs at will~\cite{moerman2022simple,korosec2021lawnmower,munoz2022catalysis}, we may be able to distinguish these mobility modes. Furthermore, separating each mode of motion should help uncover precisely the origin of subdiffusive motion~\cite{xu2011subdiffusion}, since some modes may be more prone to memory effects than others~\cite{wang2020non}. In addition, it is an open riddle to understand how these collective long-time memory effects emerge from single DNA strands, that may, as they diffuse as single molecules at a liquid-solid interface, exhibit intermittent hopping~\cite{skaug2013intermittent,suelzle2023label}.  
Such detailed microscopic inquiries will further pave the way toward advanced and rational programming of ligand-receptor motion.

\section*{Acknowledgements}

%These authors contributed equally: J.Z. and S.M.

We acknowledge fruitful discussions with Fan Cui, Florian Rehfeldt and Matthias Weiss. J.A.Z. and D.J.P. were supported by the US Department of Energy under grant DE-SC0007991 for the design and implementation of the experiments. M.H.C. acknowledges support from the Alfred P. Sloan Foundation, and from the Natural Sciences and Engineering Research Council of Canada (NSERC), RGPIN-2023-04449 / 
Cette recherche a été financée par le Conseil de recherches en sciences naturelles et en génie du Canada (CRSNG).
S.M. received funding from the European Union’s Horizon 2020 research and innovation program under the Marie Skłodowska-Curie grant agreement 839225, MolecularControl.

\section*{Data availability}

All data needed to evaluate the conclusions in the paper are present in the paper and/or the Supplementary Materials. All other data are available upon reasonable request to the authors.

\section*{Code availability}

Integration codes used to generate model curves are published~\cite{smarbach_2022_6798435} and available at \url{https://github.com/smarbach/DNACoatedColloidsInteractions}.

\section*{Competing interests}

The authors declare no competing interests.

%\newpage

%\nocite{*}

%\bibliography{DNA}% Produces the bibliography via BibTeX.


\end{document}